\documentclass{article}
\usepackage{spconf}
\usepackage{amsmath,graphicx}
\usepackage{comment}
\usepackage{amsthm, amssymb, tikz, ifthen, xspace, bm, ulem}
\usepackage[hidelinks]{hyperref}
\usepackage{dsfont}
\usepackage{booktabs}
\usepackage{subfig}
\usepackage[demo,export]{adjustbox}
\usepackage{mathtools}
\usepackage{multirow}

\newcommand\blfootnote[1]{%
  \begingroup
  \renewcommand\thefootnote{}\footnote{#1}%
  \addtocounter{footnote}{-1}%
  \endgroup
}

\newcommand{\R}{\mathbb{R}}

\theoremstyle{plain}

\newtheorem*{theorem*}{Theorem}

\theoremstyle{definition}

\input{my_symbol.sty}

\title{A Spectral Analysis of Graph Neural Networks on Dense and Sparse Graphs}
\name{Luana Ruiz, Ningyuan (Teresa) Huang, Soledad Villar}
\address{}

\begin{document}
\ninept
\maketitle
\begin{abstract}
In this work we propose a random graph model that can produce graphs at different levels of sparsity. We analyze how sparsity affects the graph spectra, and thus the performance of graph neural networks (GNNs) in node classification on dense and sparse graphs. We compare GNNs with spectral methods known to provide consistent estimators for community detection on dense graphs, a closely related task. We show that GNNs can outperform spectral methods on sparse graphs, and illustrate these results with numerical examples on both synthetic and real graphs.
\end{abstract}
%
\section{Introduction} \label{sec:intro}

\blfootnote{
LR, NH and SV are with the AMS Dept.\ and MINDS at Johns Hopkins University. SV is partially supported by ONR N00014-22-1-2126, NSF CISE 2212457, an AI2AI Amazon research award, and the NSF–Simons Research Collaboration on the Mathematical and Scientific Foundations of Deep Learning (MoDL) (NSF DMS 2031985).
}Graph neural networks (GNNs) have achieved impressive results in network data, with successful applications including genomic sequencing \cite{vrcek2023reconstruction} and satellite navigation \cite{dolan2023satellite}. Their empirical success is supported by a growing body of work on the mathematical properties of these models, such as their universality \cite{keriven2019universal}, expressive power \cite{xu2018how,boker2023fine}, stability \cite{gama19-stability} and transferability \cite{ruiz2021transferability}. GNNs are built as sequences of layers in which each layer composes a graph convolutional filterbank and a pointwise nonlinearity. A variety of constructions exist in the literature, but most of them are expressible as deep convolutional models; see \cite[Sec. I.A]{ruiz2020gnns}.

In this paper, we propose a study of GNNs on node classification tasks. Such tasks are standard graph signal processing problems where the graph is the support of the data and the node features and labels are the input and output signals respectively. 
We introduce a novel random graph model (Def. 2) encompassing both dense and relatively sparse graphs and, leveraging the fact that GNNs are spectral operators, analyze their performance on node classification in the spectral domain. We further compare GNNs against spectral embedding (SE), 
a class of established statistical methods for community detection on graphs \cite{athreya2017statistical, abbe2017community}.

Our results show that, while SEs degrade with graph sparsity (Thm. 1), under mild assumptions on the graph and on the signals there exist GNNs which perform consistently well on sparse graphs (Thm. 2). These findings are demonstrated empirically through numerical experiments on both synthetic and real-world graphs (Sec. 4).~Compared with SEs, GNNs achieve similar performance on dense graphs, and better performance on sparse graphs. 

\noindent \textbf{Related work}.~Many works have highlighted the advantages and limitations of GNNs in node classification and community detection. \cite{baranwal2021graph,baranwal2022effects} prove that the graph convolution extends the regime in which the classes are separable when the data is Gaussian. Similarly to our Thm. 2, \cite{wang2022powerful} derive conditions on the data and the graph Laplacian under which spectral GNNs are universal. \cite{huang2020combining} showed empirically that in certain node classification datasets GNNs perform no better than label propagation and spectral positional encodings. In a more recent line of work, \cite{zhu2020beyond,ma2021homophily,chen2022does} analyze how the graph homophily influence the performance of GNNs on node classification. To our knowledge, our work is the first to analyze GNN on node classification by making the distinction between dense and sparse graphs; we achieve so by proposing a general random graph model that encompasses the Stochastic Block Model (SBM) commonly used in existing works.

\section{Preliminary Definitions} \label{sec:prelim}

A graph $\bbG$ is a triplet $\bbG=(\ccalV,\ccalE,\ccalW)$ where $\ccalV=\{1,\ldots,N\}$ is the node set, $\ccalE \subseteq \ccalV \times \ccalV$ the edge set, and $\ccalW: \ccalE \to \reals$ a function assigning edge weights. We focus on unweighted, undirected and connected graphs $\bbG$, so that $\ccalW:\ccalE \to \{0,1\}$, $\ccalW(i,j)=\ccalW(j,i)$ for all $i,j$ and there is a single connected component. We represent the graph $\bbG$ by its adjacency matrix $\bbA \in \reals^{N \times N}$, defined as $[\bbA]_{ij} = \ccalW(i,j)$ if $(i,j) \in \ccalE$ and $0$ otherwise. Since $\bbA$ is symmetric, it can be diagonalized as $\bbA = \bbV\bbLam\bbV^\top$. The diagonal elements of $\bbLam$ are the eigenvalues $\lambda_i \in \R$, $|\lambda_1 | \geq \ldots \geq |\lambda_N|$, and the columns of $\bbV$ the corresponding eigenvectors $\bbv_i$, $1 \leq i \leq N$.  

We assume that the nodes of $\bbG$ can carry data, which is represented in the form of \textit{graph signals} \cite{ortega2018graph,sandryhaila13-dspg}. A graph signal is a vector $\bbx \in \reals^N$ where $[\bbx]_i$ is the value of the signal of the node $i$. More generally, graphs can also carry $D$-dimensional signals $\bbX \in \reals^{N \times D}$, where each column of $\bbX$, denoted $\bbx^d$, is a \textit{node feature}.

Closely related to node classification, community detection consists of clustering nodes $i \in \ccalV$ into $K$ communities. 
The goal of community detection is thus to obtain a graph signal $\bbY \in [0,1]^{N \times K}$ where each row $[\bbY]_{i\cdot}$ represents the \textit{community assignment} of node $i$ (potentially overlapping \cite{xie2013overlapping}). In this paper, we assume non-overlapping communities, so that $[\bbY]_{i\cdot}=\texttt{one-hot}(k)$ (i.e., $[\bbY]_{ij}=1$ for $j=k$ and $0$ for $j \neq k$) implies that node $i$ is in community $k$. 

There are many variants of community detection \cite{abbe2017community}. For example, the number of communities $K$ may or may not be predefined \cite{choi2012stochastic}, and the problem can be solved in an unsupervised or supervised manner \cite{cai2020weighted}. In this paper, we assume that $K$ is given and solve the problem with supervision.
Formally, given a graph $\bbG$ and a signal $\bbX$, and a true community assignment matrix $\bbY$, we fix a training set consisting of a subset $\ccalT = \{i_1, \ldots, i_M\} \subset \ccalV$ of the graph nodes. This training set is used to define a node selection matrix $\bbM_\ccalT \in \{0,1\}^{M\times N}$ where $[\bbM_\ccalT]_{ij}=1$ only for $i=m$, $j=i_m$, and the masked input signal $\bbX_\ccalT \in \reals^{N \times D}$ where $[\bbX_\ccalT]_{i \cdot} = [\bbX]_{i \cdot}$ for $i \in \ccalT$ and $0$ otherwise. We then use $\ccalT$ to solve the following optimization problem
\begin{equation} \label{eqn:erm}
    \min_f \ell(\bbM_\ccalT\bbY,\bbM_\ccalT f(\bbA,\bbX_\ccalT))
\end{equation}
where $\ell:\reals^{M\times K} \times \reals^{M \times K} \to \reals$ is a loss and $f: \reals^{N \times N} \times \reals^{N \times D} \to \reals^{N\times K}$ is a parametric function.

Typically, the function $f$ is parametrized as 
\begin{equation} \label{eqn:parametrization}
    f = c \circ \phi
\end{equation}
where $c$ is a classifier, $\phi$ is an embedding, and $\circ$ denotes function composition. We will consider the case where the embedding is obtained via SEs in Sec. \ref{sbs:sbm}, and via GNNs in Sec. \ref{sbs:gnns}.

\subsection{Stochastic Block Model and Spectral Embeddings} \label{sbs:sbm}

The canonical statistical model for graphs with communities is the stochastic block model (SBM).

\begin{definition}[Stochastic Block Model] \label{defn:sbm}
A SBM graph with $K$ communities is defined as a graph $\bbG$ with adjacency matrix $\bbA \in \{0,1\}^{N \times N}$ given by
\begin{equation*}
\bbA \sim \mbox{Ber}(\bbP),\ \, 
\bbP = \bbY \bbB \bbY^{\top}
\label{eqn:SBM}
\end{equation*}
where $\bbY \in \{0,1\}^{N \times K}$ is the community assignment matrix $\bbY_{i\cdot} = \texttt{one-hot}(k)$, and $\bbB \in [0,1]^{K \times K}$ is a full-rank matrix representing the block connection probability. 
\end{definition}

Spectral methods for community detection are inspired by the spectral decomposition of the SBM. Consider for example the case where $K=2$, $\bbB=[p\ q; q\ p]$, $p \neq q$, and the communities are balanced, i.e., $N$ is even and both communities have size $N/2$. Relabeling $\ccalV$ so that the first $N/2$ nodes belong to the first community and the remaining $N/2$ to the second, we see that the eigenvectors of $\mbE\bbA \equiv \bbP$, the expected adjacency, are given by
\begin{equation} \label{eqn:sbm_eig}
    [\bbv_1(\mbE\bbA)]_i = \frac{1}{\sqrt{N}},\ [\bbv_2(\mbE\bbA)]_i = 
    \begin{cases}
    -1/\sqrt{N}, \ i \leq N/2 \\
    +1/\sqrt{N}, \ i > N/2 \text{.}
    \end{cases}
\end{equation}
For a graph $\bbG$ sampled from this model, with sufficiently large $N$ and mild assumptions on $p,q$, we can thus expect the eigenvector $\bbv_2(\bbA)$ to provide a good estimate of its community structure, i.e., $\bbv_k(\bbA) \approx \bbv_k(\mbE\bbA), k \in \{1,2 \}$.  

Real-world graphs $\bbG$ have more intricate sparsity patterns than the SBM, but it is reasonable to assume that if the graph $\bbG$ has two balanced communities, for some permutation of the nodes its adjacency matrix $\bbA$ can be approximately written as $\bbA = \bbA_{\tiny \mbox{SBM}} + \bbE$, where $\bbA{\tiny \mbox{SBM}}$ is as in Def. \ref{defn:sbm} and $\bbE$ can be seen as a perturbation satisfying $\|\bbE\|_2 < \|\bbA{\tiny \mbox{SBM}}\|_2$. As such, the first two eigenvectors of $\bbA$ still ``embed'' community information. More generally, in graphs $\bbG$ with $K>2$ balanced communities, the community information is ``embedded'' in the first $K$ eigenvectors. Based on this observation, the order-$K$ \textit{spectral embedding} of a graph $\bbG$ is defined as
\begin{equation} \label{eqn:spectral_embedding} 
    \phi_{\tiny \mbox{SE}}(\bbA) = [\bbv_1\ \bbv_2\ \ldots\ \bbv_{K-1}\ \bbv_K] = \bbV_K,
\end{equation}
i.e., as the concatenation of the first $K$ eigenvectors of $\bbA$. Variants of SE tailored for sparse graphs propose replacing $\bbA$ with other graph operators, such as the normalized adjacency matrix $\tilde{\bbA}  \coloneqq \bbD^{-0.5} \bbA \bbD^{-0.5}$ where $\bbD$ is the degree matrix \cite{cape2019spectral}, the non-backtracking operator \cite{abbe2017community}, etc. 

Note that $\phi_{\tiny \mbox{SE}}$ is nonparametric; it can be obtained directly from the graph without node label supervision. When we use spectral embeddings in \eqref{eqn:parametrization}, the only parameters that are learned are those of the classifier $c$. E.g., choosing a linear classifier yields a simple parameterization of $f$ as
$f(\bbA) = c \circ \phi_{\tiny \mbox{SE}}(\bbA) = \mbox{\texttt{softmax}}(\bbV_K\bbC)$
where $\bbC \in \reals^{K \times K}$ is learned. More generally, it is possible to use embeddings $\phi_{\tiny \mbox{SE}}(\bbA) = \bbV_{\tilde{K}}$ with $\tilde{K}>K$, i.e., with a larger number of eigenvectors than that of communities, in which case $\bbC \in \reals^{\tilde{K}\times K}$.

An important observation to make is that $ \phi_{\tiny \mbox{SE}}$ (and so $f$) do not need to depend on $\bbX$, but if such node features are available, they can be incorporated into the spectral embedding in different ways, e.g., \cite{yang2013node, binkiewicz2017covariate, arroyo2021inference, mu2022spectral, Mele2022discrete}. We consider an approach similar to \cite{arroyo2021inference}, by first embedding the node feature covariance and concatenating it with the spectral embedding. More precisely, let $\bbV'_{\kappa}$ be the first $\kappa$ eigenvectors of the covariance matrix $\bbX \bbX^{\top}$, then the \textit{feature-aware} spectral embedding is defined as 
\begin{equation}
  \label{eqn:node_sp_emb}
    \phi_{\tiny \mbox{SE}}(\bbA;\bbX) = [\bbV_K\ \ \bbV'_{\kappa}].
\end{equation}

\subsection{Graph Neural Networks} \label{sbs:gnns}

Given a graph $\bbG$ with adjacency matrix $\bbA \in \reals^{N \times N}$ and a graph signal $\bbx \in \reals^{N}$, a graph convolution (or filter) is given by \cite{du2018graph}
\begin{equation} \label{eqn:graph_convolution}
   \bbu = \sum_{k=0}^{K-1} h_k \bbA^k \bbx
\end{equation}
where $h_0, \ldots, h_{K-1}$ are the filter coefficients or taps. More generally, if $\bbX \in \reals^{N \times D}$ and $\bbU \in \reals^{N\times G}$ have $D$ and $G$ features respectively, we write
\begin{equation} \label{eqn:graph_convolution_multifeature}
   \bbU = \sum_{k=0}^{K-1} \bbA^k \bbX \bbH_k
\end{equation} 
where the filter parameters are now collected in the matrices $\bbH_0, \ldots,$ $\bbH_{K-1} \in \reals^{D \times G}$.

Graph neural networks (GNNs) are deep convolutional architectures where each layer composes a graph convolution \eqref{eqn:graph_convolution_multifeature} and a pointwise nonlinearity $[\sigma(\bbU)]_{ij}=\sigma([\bbU]_{ij})$, e.g., the ReLU or the sigmoid. The $\ell$th layer of a GNN can thus be written as
\begin{equation} \label{eqn:gnn}
    \bbX_\ell = \sigma \left( \sum_{k=0}^{K-1} \bbA^k \bbX_{\ell-1} \bbH_{\ell k} \right)
\end{equation}
where $\bbX_{\ell-1} \in \reals^{N \times F_{\ell-1}}$ and $\bbX_{\ell} \in \reals^{N \times F_{\ell}}$ are the input and output to this layer with $F_{\ell-1}$ and $F_\ell$ features each. If the GNN has $L$ layers, its input and output are $\bbX_0 = \bbX \in \reals^{N \times F_0}$ and $\bbX_L \in \reals^{N \times F_L}$.

The GNN in \eqref{eqn:gnn} may be used to parametrize $\phi$ in \eqref{eqn:parametrization}, in which case we define the \textit{GNN embedding}
\begin{equation} \label{eqn:gnn_embedding}
    \phi_{\tiny \mbox{GNN}}(\bbA,\bbX)=\bbX_L \text{.}
\end{equation}
Note that, unlike the spectral embedding \eqref{eqn:spectral_embedding}, \eqref{eqn:gnn_embedding} is parametric on $\{\bbH_{\ell k}\}_{\ell,k}$ and always needs an input signal $\bbX$ (if an input signal is not available, $\bbX$ may be a random signal, for example). A typical parametrization of $f$ for GNN embeddings is
$f(\bbA,\bbX) = c \circ \phi_{\tiny \mbox{GNN}}(\bbA, \bbX) = \mbox{\texttt{softmax}}(\bbX_L\bbC)$
where $\bbC \in \reals^{F_L \times K}$ is a linear classifier over $F_L$ node features. This is equivalent to a $L+1$-layer GNN with $K=1$ and softmax nonlinearity in the last layer.


\section{Main Results} \label{sec:main}

In the following, we introduce a random graph model for both dense and sparse graphs. We use this model to prove a result that helps explain the limitations of spectral embeddings on sparse graphs. We then show that under mild assumptions on both the graph and the input signal, GNNs give access to entire spectrum, and thus can learn embeddings that are more expressive than spectral embeddings.

\subsection{A Graph Model for Dense and Sparse Graphs} \label{sbs:graph_model}

Def. \ref{def:graphex} introduces a random graph model allowing to model graphs with varying levels of sparsity according to a sparsity parameter $\gamma$.

\begin{definition}[Dense-Sparse Graph Model (DSGM)] \label{def:graphex}
A DSGM graph with kernel $\bbW$ and sparsity parameter $\gamma$ is defined as a graph $\bbG$ with adjacency matrix $\bbA \in \{0,1\}^{N \times N}$ given by
\begin{equation*}
    [\bbA]_{ij} = [\bbA]_{ji} \sim \mbox{Ber}(\bbW(u_i,u_j)),\   
    u_{i} = 
    \begin{cases}
    u_{i-1} + \gamma, \ 2 \leq i \leq N \\
    -\lfloor\dfrac{n}{2}\rfloor\gamma + \dfrac{\gamma}{2},\ i=1
    \end{cases} \label{eqn:ui}
\end{equation*}
where $\bbW: \reals^2 \to [0,1]$ is symmetric, $\|\bbW\|_{L^2} < \infty$, and $\gamma > 0$.
\end{definition}

This model allows sampling both dense and sparse graphs because, since $\bbW$ has vanishing tails (or can be mapped to a kernel that does by some measure-preserving transformation), for a fixed $N$ the graph is sparser for larger $\gamma$. 

The kernel $\bbW$ defines a self-adjoint Hilbert Schmidt operator. Hence, it has a real spectrum given by
\begin{equation}
    \int_{-\infty}^{\infty} \bbW(u,v) \varphi_i(u)du = \lambda_i \varphi_i(v)
\end{equation}
where the eigenvalues $\lambda_i$ are countable and the eigenfunctions $\varphi_i$ form an orthonormal basis of $L^2$. 
By convention, the eigenvalues are ordered as $|\lambda_1|\geq|\lambda_2|\geq\ldots$. Moreover, $|\lambda_i| \leq \infty$ for all $i$, and $\lambda_i \to 0$ as $i \to \infty$ with zero being the only accumulation point.

We further introduce the notion of a kernel induced by a graph, which will be useful in future derivations. For $N\geq 2$, the kernel induced by the graph $\bbG_N$ with adjacency $\bbA_N$ and sparsity parameter $\gamma$ is defined as
\begin{equation} \label{eqn:induced}
\bbW_{N}(u,v)=\sum_{i=1}^{N-1}\sum_{j=1}^{N-1} [\bbA_N]_{ij}\mbI(u \in I_i)\mbI(v \in I_j) 
\end{equation}
where $I_i=[u_i,u_{i+1})$ for $1\leq i\leq N-2$, $I_{N-1} = [u_{N-1},u_N]$, and $u_i$ is as in Def. \ref{eqn:ui}.

\subsection{Limitations of Spectral Embeddings} \label{sbs:limitations}

To discuss community detection on graphs sampled from a DSGM (Def. \ref{def:graphex}), we assume that the kernel $\bbW$ exhibits community structure. 
For simplicity, we focus on $2$ communities but the discussions can be easily extended to $K$ communities. 
Inspired by the degree-corrected SBM \cite{karrer2011stochastic, qin2013regularized}, in Def. \ref{eqn:dc_sbm} we introduce the degree-corrected stochastic block kernel (SBK) as the canonical kernel for DSGMs with $2$ balanced communities. This model is suitable to model sparse graphs and well-studied in the spectral embedding literature \cite{qin2013regularized, cape2019spectral}. To ensure that models based on these kernels are valid DSGMs, we restrict attention to finite-energy degree functions $\theta$.

\begin{definition}[Degree-Corrected SBK] \label{eqn:dc_sbm} 
The degree-corrected SBK with $2$ communities is given by
\begin{equation*}
 \bbW(u,v) = 
    \begin{cases}
    \theta(u) \, \theta(v) \, p ,\quad u v \ge 0 \\
     \theta(u) \, \theta(v) \, q,\quad u v < 0 
    \end{cases}
\end{equation*}
where $\theta:\reals\to[0,1]$, $\theta \in L^2$, is the degree function. The true community assignment is $Y(u) =  [1\ 0]\mbI(u \ge 0) + [0\ 1]\mbI(u < 0)$, which is independent of $\theta$.
\end{definition}

It is not difficult to see that the first $2$ eigenfunctions of $\bbW$ in Def. \ref{eqn:dc_sbm} reveal the community structure \footnote{$\varphi_1(u) = \theta(u)/C, \varphi_2(u) =  \big( -\theta(u) \mbI(u < 0) + \theta(u) \mbI(u \ge 0) \big)/C, \\ \text{ where } C \coloneqq \int \theta(u) du$.}. 
For graphs $\bbG_N$ sampled as in Def. \ref{def:graphex} from the DSGM with degree-corrected kernel as in Def. \ref{eqn:dc_sbm}, the true community assignment is given by $[\bbY]_{i\cdot} = Y(u_i)$ for $1 \leq i \leq N$. As such, the quality of the estimate of the community assignment given by the first $2$ (or, more generally, the first $K$) eigenvectors of $\bbG_N$ will depend on both (i) how close the eigenvalues $\lambda_{k}(\bbG_N)$ are to the kernel eigenvalues $\lambda_{k}(\bbW)$ (as this can affect their ordering) and (ii) how close the eigenvectors $\bbv_{k}$ are to the eigenfunctions $\varphi_{k}$. These differences are upper bounded by Thm. \ref{thm:comm_conc}.

\begin{theorem}[Eigenvalue and eigenvector concentration] \label{thm:comm_conc}
Let $\bbG_N$ be a graph sampled from the DSGM in Def. \ref{def:graphex}, where $N$ satisfies \cite[Ass. AS4]{ruiz2021transferability}. Let $c \leq \lfloor N/2\rfloor\gamma-\gamma/2$, and assume that:
\begin{enumerate}
    \item $\bbW$ is $A_w$-Lipschitz in $[-c,c] \times [-c,c]$ (see \cite[Ass. AS2]{ruiz2021transferability})
    \item $\int_{|v|\geq c} \int_{|u|\geq c} \bbW(u,v)dudv < \epsilon(c)$.
\end{enumerate}
Then, with probability at least $1-\chi$, the difference between the $k$th eigenvalue of $\bbG_N$ and $\bbW$, $1\leq k\leq K$, is bounded by
\begin{align*}
\begin{split}
    |\lambda_k(\bbW_N)-\lambda_k(\bbW)| &\leq 4 A_w c \gamma + {\beta(\chi,N)}{N^{-1}} + \epsilon(c) \\
    &\leq 2 A_w N \gamma ^2 + {\beta(\chi,N)}{N^{-1}} + \epsilon(c)
\end{split}
\end{align*}
and the difference between their $k$th eigenvectors by
\begin{align*}
\begin{split}
    \|\varphi_k(\bbW_N)-\varphi_k(\bbW)\| 
    \leq \frac{\pi}{2\delta_k} \bigg(4A_w c \gamma
    + {\beta(\chi,N)}{N^{-1}} + \epsilon(c)\bigg)
\end{split}
\end{align*}
where $\bbW_N$ is the kernel induced by $\bbG_N$ \eqref{eqn:induced}\footnote{See \cite[Lemma 2]{ruiz2020graphonsp} for the relationship between $\lambda_k(\bbG_N)$, $\bbv_k(\bbG_N)$ and $\lambda_k(\bbW_N)$, $\varphi_k(\bbW_N)$.}, $\delta_k = \min_i{\{|\lambda_k(\bbW)-}$ ${\lambda_i(\bbW_N)|,|\lambda_k(\bbW_N)-\lambda_i(\bbW)|\}}$ and $\beta(\chi,N)$ is sublinear in $N$ and as in \cite[Def. 7]{ruiz2021transferability}.
\end{theorem}
\begin{proof}
Refer to the extended version in this \href{https://arxiv.org/pdf/2211.03231.pdf}{link}.
\end{proof}

\begin{figure}[t]
  \centering
\subfloat[$\bbW$ ]{\includegraphics[height=1.9cm,valign=c]{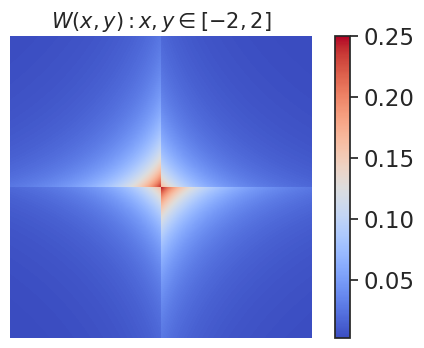}}
\qquad
\subfloat[Dense $\bbG_N^d$]{\includegraphics[height=1.9cm,valign=c]{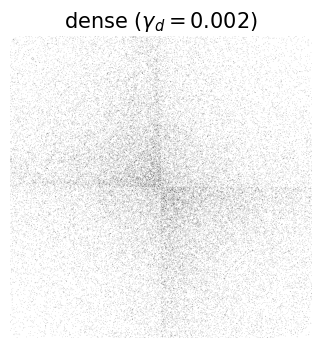}  }
\qquad
\subfloat[Sparse $\bbG_N^s$]{\includegraphics[height=1.9cm,valign=c]{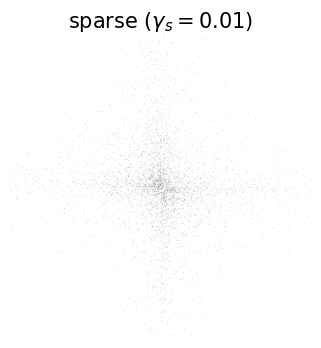}  }
\caption{Kernel $\bbW:\R^2 \to [0,1]$ visualized in $[-2,2]^2$ and sampled graphs with different sparsity levels $\gamma$.}
  \label{fig:graphex_model}
\end{figure}

This theorem shows that the differences between the eigenvalues and eigenvectors of the graph and the underlying random graph model are upper bounded by terms that increase with $\gamma$. Consider a dense graph $\bbG_N^d$ and a sparse graph $\bbG_N^s$ sampled from DSGMs with same kernel $\bbW$ but different sparsity parameters $\gamma_d \ll \gamma_s$. If $N$ and $c$ are large enough for the term depending on $4A_w c \gamma$ to dominate the bound in the dense case, the bound on the difference between eigenvalues and eigenvectors in the sparse case is much larger than in the dense case.
In the context of community detection, this can be interpreted to mean that, since $\varphi_k(\bbW_N^d)$ is close to $\varphi_k(\bbW)$ for dense graphs, some linear combination of the eigenvectors $\bbv_k(\bbG_N^d)$ provides a good estimate of the true community assignment $\bbY$.
This is not true for the eigenvectors $\bbv_k(\bbG_N^s)$ of the sparse graph, since $\varphi_k(\bbW_N^s)$ is further away from $\varphi_k(\bbW)$.
Another way to think about this is that on dense graphs most of the ``community information'' is on the first $K$ eigenvectors. On sparse graphs, it is more spread throughout the spectrum. This implies that while spectral embeddings may be effective for community detection on dense graphs, they are less likely to be effective in sparse graphs. We further demonstrate this empirically in Sec. \ref{sec:exp}.

\subsection{Graph Neural Networks for Community Detection} \label{sbs:gnns_comm}

In sparse graphs, GNN embeddings are a better option than spectral embeddings because, provided that the input signal $\bbX_{\ccalT}$ in \eqref{eqn:erm} is not orthogonal to any of the graph's eigenvectors, GNNs ``have access'' to the entire spectrum. Moreover, if the true community assignment signal is $\bbY$, a GNN can always represent $\bbY$ with $K\leq N$ in \eqref{eqn:gnn}. These claims are formally stated for the simple graph convolution \eqref{eqn:graph_convolution} in Thm. \ref{thm:power}. They can be readily extended to multi-feature graph convolutions \eqref{eqn:graph_convolution_multifeature} and GNNs \eqref{eqn:gnn} where the nonlinearity $\sigma$ preserves the sign (e.g., the hyperbolic tangent).

\begin{theorem}[Expressive power of graph convolution] \label{thm:power}
Let $\bbG$ be a symmetric graph with full-rank adjacency matrix $\bbA \in \reals^{N\times N}$ diagonalizable as $\bbA=\bbV\bbLam\bbV^\top$ where all eigenvalues have multiplicity one. Let $\bbx \in \reals^N$ be an input signal satisfying $[\bbV^\top\bbx]_i \neq 0$ for $1\leq i\leq N$. 
Consider the graph convolution $\hby=\sum_{k=0}^{K-1}h_k \bbA^k \bbx$ \eqref{eqn:graph_convolution}. Then, the following hold:
\begin{enumerate}
    \item For all $K\geq 1$, there exist $h_0, \ldots, h_{K-1} \in \reals$ such that $\hby$ satisfies $[\bbV^\top\hby]_i \neq 0$ for every $i$.
    \item Let $\bby \in \reals^N$ be a target signal. There exist $K\leq N$ coefficients $h_0, \ldots, h_{K-1} \in \reals$ for which $\hby$ satisfies $\hby=\bby$.
\end{enumerate}
\end{theorem}
\begin{proof}
Refer to the extended version in this \href{https://arxiv.org/pdf/2211.03231.pdf}{link}. Also note that this theorem is analogous to \cite[Thm. 4.1.]{wang2022powerful}, which proves a similar result for GNNs based on the graph Laplacian.
\end{proof}

Note that the assumptions of Thm. \ref{thm:power} are not too restrictive; most real-world graphs are full rank, and even a random signal $\bbx \in \reals^N$---which may be used as the input in \eqref{eqn:gnn_embedding} when $\bbx$ is not given---satisfies $[\bbV^\top\bbx]_i\neq 0$ with high probability. It is also worth pointing out that while $K \leq N$ is necessary to \textit{exactly represent} $\bby$, in practice small $K$ is often enough to obtain good \textit{approximations} of the true community assignment as illustrated in Sec. \ref{sec:exp}. This is another reason why in practical, large graph settings, GNN embeddings are advantageous w.r.t. spectral embeddings: a small number of matrix-vector multiplications requires less computations than calculating a number of eigenvectors at least as large as the number of communities. 

\section{Experiments} \label{sec:exp}
 
\begin{figure}[t]
  \centering
\includegraphics[width=0.5\textwidth]{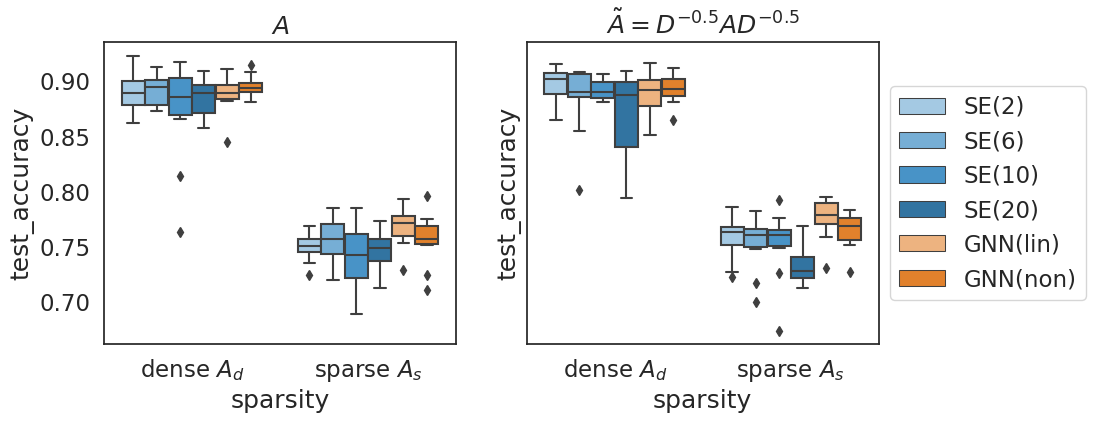}
  \caption{Test accuracy for different sparsity levels of the sampled graphs. GNNs perform better than SEs in sparse graphs for both operators $\bbA, \tilde{\bbA}$.}
  \label{fig:graphex_exp}
\end{figure}
 
In what follows, we conduct simulations on synthetic graphs sampled from a DSGM (Section \ref{sec:exp-dsgm}) and real-world graphs (Section \ref{sec:exp-real}). For completeness, we consider graph operators $\bbA, \tilde{\bbA}$. Our empirical results validate our theoretical analysis and show that GNNs outperform spectral embedding for community detection in sparse graphs.\footnote{All the simulations and code are available in this \href{https://arxiv.org/pdf/2211.03231.pdf}{link}.}

\subsection{Experiments on Synthetic Graphs} \label{sec:exp-dsgm}

\noindent \textbf{Setup.} We consider the following kernel
\begin{equation}
\bbW(u,v) = \begin{cases}
\frac{p}{(|u|+1)^{2}(|v|+1)^{2}} & uv \ge  0\\
\frac{q}{(|u|+1)^{2}(|v|+1)^{2}} & uv < 0. \label{eqn:a_graphex}
\end{cases}
\end{equation}
The graphs $\bbG$ are sampled from the DSGM with kernel $\bbW$ above following Def. \ref{def:graphex}, with $N=1000$ and different choices of density parameter $\gamma_d = 0.002, \gamma_s = 0.01$ as illustrated in Fig. \ref{fig:graphex_model}. The node features $\bbX$ are sampled from a mixture of two Gaussians in $\R^2$ where $\bbmu_0 = - \bbmu_1 = [1, 1], \bbSigma_0 = \bbSigma_1 = \bbI/4$.
For each tuple $(\bbG,\bbX)$, we randomly split the nodes in each community by $50/50$ to create the training and test sets. We compare spectral embeddings with various choices of $K$ against GNNs. 

\begin{figure}[t]
  \centering
\subfloat[$\hby$ (linear) for $\bbG_N^d$ ]{\includegraphics[width=0.2\textwidth,height=2cm,valign=c]{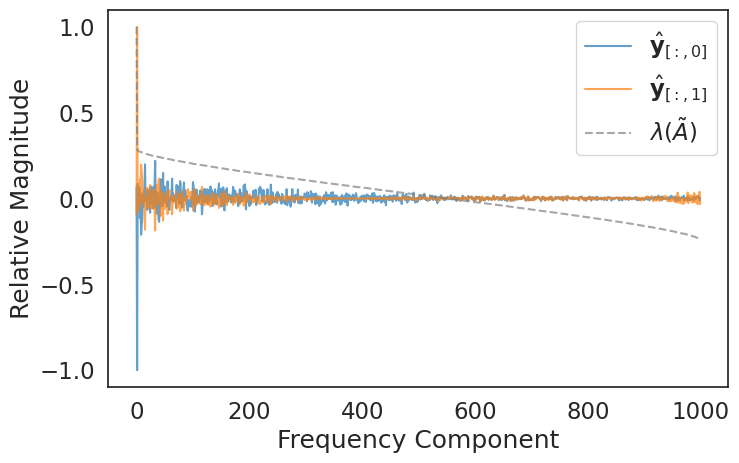}}
\qquad
\subfloat[$\hby$ (linear) for $\bbG_N^s$]{\includegraphics[width=0.2\textwidth,height=2cm,valign=c]{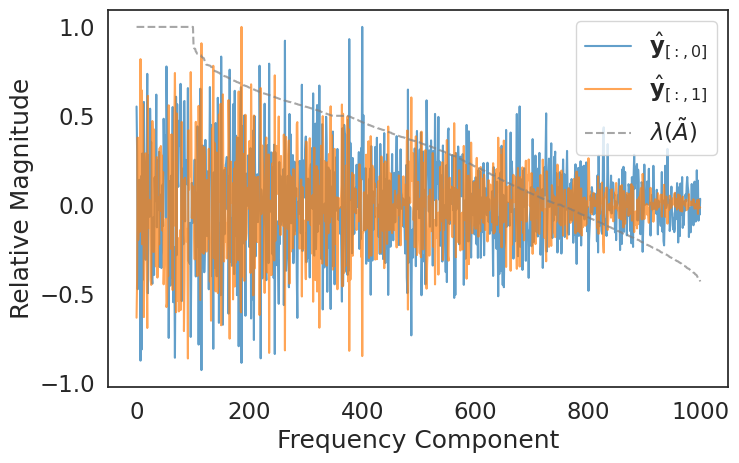}  }
\qquad
\subfloat[$\hby$ (nonlinear) for $\bbG_N^d$ ]{\includegraphics[width=0.2\textwidth,height=2cm,valign=c]{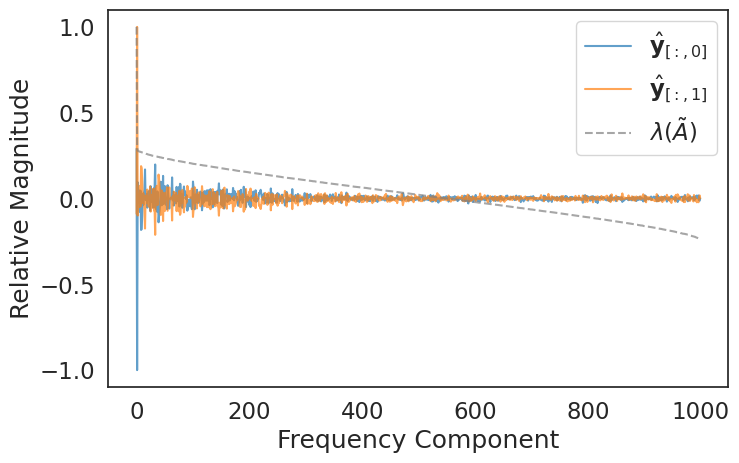}}
\qquad
\subfloat[$\hby$ (nonlinear) for $\bbG_N^s$]{\includegraphics[width=0.2\textwidth,height=2cm,valign=c]{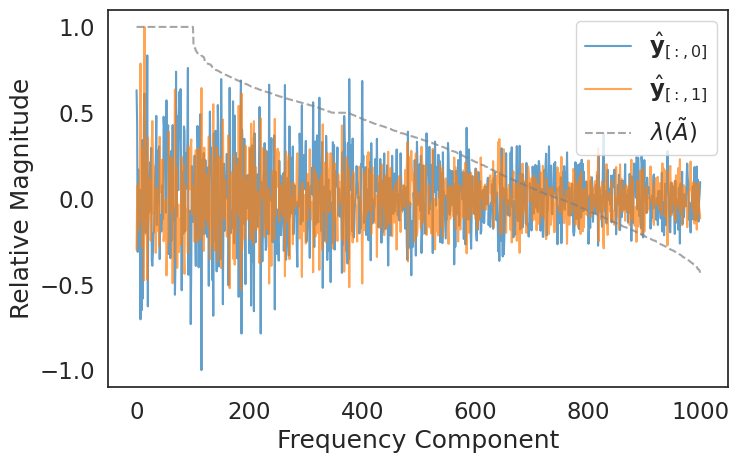}  }
\caption{Frequency responses $\hby$ of GNNs using $\tilde{\bbA}$ on $\bbG_N^d$ and $\bbG_N^s$. In the dense case (left), although the optimal frequency response is a step-function on the first two components, GNNs spread energies on the remaining components, adding noise; In the sparse case (right), the community information spreads widely across the spectrum and thus GNNs outperform spectral embedding. Nonlinear GNNs (bottom) leverage the spectrum more uniformly than linear convolutions (top). Eigenvalues of $\tilde{\bbA}$ (dashed) are sorted in decreasing order.}
  \label{fig:freq}
\end{figure}

\noindent \textbf{Results.} Fig. \ref{fig:graphex_exp} shows that spectral embedding with $K=2$ outperforms GNNs in dense graphs while GNNs are more competitive in sparse graphs. Fig. \ref{fig:freq} depicts the frequency response $\hby$ from the trained GNN model using $\tilde{\bbA}$: (a) shows that, in the dense graph, GNNs indeed attend to frequency components other than the first two eigenvectors, which increase the noise/variance of the embedding and thus degrades the downstream classification performance, confirming the discussion in Thm. \ref{thm:power}; (b) shows that, in the sparse graph, GNNs increasingly attend to higher-frequency components, which are useful since they may also encode community information; spectral embeddings exhibit higher variance, and can benefit from choosing suitably larger embedding dimension.

\subsection{Experiments on Real-World Graphs} \label{sec:exp-real}

\noindent \textbf{Setup.} We consider the Wikipedia webpage network Chameleon, a heterophilous benchmark graph with $5$ communities introduced in \cite{rozemberczki2021multi}. We treat the original Chameleon network ($|\ccalV|=2277$, average degree $13.8$) as the dense baseline, and randomly drop a fraction of its edges to obtain the sparse(r) graphs. We then evaluate GNNs and spectral embedding in the original and sparsified graphs. For each sparsity level, we randomly generate $10$ sparsified graphs. 

\noindent
\textbf{Results.} Table~\ref{tab:exp_real} shows that GNNs and spectral embeddings both perform well in the original graph. Yet, in the sparsified graphs (``Drop(20)'', ``Drop(70)''), performance degradation in GNNs is smaller than spectral embeddings. 
Moreover, in sparsified graphs, spectral embeddings with large $K$ are numerically unstable and computationally intensive due to the presence of many small eigenvalues. These findings show that GNNs can detect communities more accurately and efficiently than spectral methods in sparse graphs.

\begin{table}[htb!]
\scriptsize
\caption{Test accuracy on Chameleon graphs, reported as \texttt{mean($\pm$stderr)} across 10 data splits and 10 sparsified subgraphs.}
\label{tab:exp_real}
\centering
\resizebox{\columnwidth}{!}{\begin{tabular}{cccccc}
\hline \hline
Graph & Operator	&	SE(150)	&	SE(200)	&	GNN(lin)	&	 GNN(non) 	\\
\hline \hline
\multirow{2}{*}{Original} & $\bbA$ 	&	\textbf{57.29 $\pm$ 0.69}	&	56.97 $\pm$ 0.59	&	56.27 $\pm$ 0.69	&	54.38 $\pm$  0.97	\\
 & $\tilde{\bbA}$		&	52.70 $\pm$ 0.36	&	53.84 $\pm$ 0.43	&	55.60 $\pm$  0.70	&	\textbf{55.90 $\pm$ 0.73  }	\\
  \hline
\multirow{2}{*}{Drop(20)} & $\bbA$ &	53.20 $\pm$ 0.21	&	53.30 $\pm$  0.22	&	\textbf{53.91 $\pm$  0.25}	&	52.69 $\pm$ 0.29	\\
	&  $\tilde{\bbA}$ &	49.42 $\pm$ 0.21	&	51.53 $\pm$  0.19		& 54.45	 $\pm$ 0.21 &	\textbf{54.66 $\pm$ 0.22 }	\\
\hline
\multirow{2}{*}{Drop(70)}  & $\bbA$	&	45.47 $\pm$ 0.20	&	45.12 $\pm$ 0.22	&	\textbf{46.21 $\pm$ 0.23}	&	45.95 $\pm$ 0.24	\\ 
	&$\tilde{\bbA}$ & 41.21 $\pm$ 0.19	&	42.51 $\pm$ 0.27	&	50.10 $\pm$	0.19 &\textbf{50.25 $\pm$ 0.21}\\
\hline \hline
\end{tabular}}
\end{table}

\label{sec:refs}
\bibliography{myIEEEabrv,refs,bib_dissertation}
\bibliographystyle{IEEEbib}

\section{Appendix} \label{sec:app}

\subsection{Proof of Theorem \ref{thm:comm_conc}}\label{sec:comm_conc}

The proof of Theorem \ref{thm:comm_conc} relies on slight variations of the Courant-Fisher and Davis-Kahan theorems, stated here as Propositions \ref{prop:eigenvalue_diff} and \ref{thm:davis_kahan}.

\begin{proposition}[Variant of Courant-Fisher] \label{prop:eigenvalue_diff}
Let $\bbW:[0,1]^2\to[0,1]$ and $\bbW':[0,1]^2\to[0,1]$ be two graphons with eigenvalues given by $\{\lambda_i(T_\bbW)\}_{i\in\mbZ\setminus\{0\}}$ and $\{\lambda_i(T_{\bbW'})\}_{i\in\mbZ\setminus\{0\}}$, ordered according to their sign and in decreasing order of absolute value, where $T_\bbW$ denotes the integral linear operator with kernel $\bbW$. Then, for all $i \in \mbZ \setminus \{0\}$, the following inequalities hold
\begin{equation*}
|\lambda_i(T_{\bbW'})-\lambda_i(T_\bbW)| \leq \vertiii{T_{\bbW'-\bbW}} \leq \|\bbW'-\bbW\|\ .
\end{equation*}
\end{proposition}
\begin{proof} \renewcommand{\qedsymbol}{}
See \cite[Proposition 4]{ruiz20-transf}.
\end{proof}

\begin{proposition}[Variant of Davis-Kahan]\label{thm:davis_kahan}
Let $T$ and $T^\prime$ be two self-adjoint operators on a separable Hilbert space $\ccalH$ whose spectra are partitioned as $\gamma \cup \Gamma$ and $\omega \cup \Omega$ respectively, with $\gamma \cap \Gamma = \emptyset$ and $\omega \cap \Omega = \emptyset$. If there exists $d > 0$ such that $\min_{x \in \gamma,\, y \in \Omega} |{x - y}| \geq d$ and $\min_{x \in \omega,\, y \in \Gamma}|{x - y}| \geq d$, then the spectral projections $E_T(\gamma)$ and $E_{T^\prime}(\omega)$ satisfy
\begin{equation*}\label{eqn:davis_kahan}
	\vertiii{E_T(\gamma) - E_{T^\prime}(\omega)} \leq \frac{\pi}{2} \frac{\vertiii{{T - T^\prime}}}{d}
\end{equation*}
\end{proposition}
\begin{proof} \renewcommand{\qedsymbol}{}
See \cite{seelmann2014notes}.
\end{proof}

We thus only need to bound $\|\bbW-\bbW_N\|$. To do so, define $\overline{\bbW}_N$ as
\begin{equation} \label{eqn:induced}
\overline{\bbW}_{N}(u,v)=\sum_{i=1}^{N-1}\sum_{j=1}^{N-1} \bbW(u_i,\bbu_j)\mbI(u \in I_i)\mbI(v \in I_j) 
\end{equation}
where $I_i=[u_i,u_{i+1})$ for $1\leq i\leq N-2$, $I_{N-1} = [u_{N-1},u_N]$, and $u_i$ is as in \eqref{eqn:ui}.
Using the triangle inequality, we can write
\begin{equation}
    \|\bbW-\bbW_N\| \leq \|\bbW-\overline{\bbW}_N\| + \|\overline{\bbW}_N-\bbW_N\| \text{.}
\end{equation}
The norm difference between $\overline{\bbW}_N$ and $\bbW_N$ is bounded as $N^{-1}\beta(\chi,N)$ by \cite[Proposition 4]{ruiz2021transferability} and by the fact that $\|\bbW_N\|_{L_2}=N^{-1}\|\bbA_N\|_2$ (see \cite[Lemma 2]{ruiz2020graphonsp}). Let us now derive a bound for $\|\bbW-\overline{\bbW}_N\|$.

By definition of the $L^2$ norm,
\begin{align} \label{eqn:integral_term}
\begin{split}
    \|\bbW-\overline{\bbW}_N\| &= \sqrt{\int_{-\infty}^\infty |\bbW(u,v)-\overline{\bbW}_N(u,v)|^2dudv} \\
                               &\leq \sqrt{\int_{|v|< c} \int_{|u|< c} |\bbW(u,v)-\overline{\bbW}_N(u,v)|^2dudv} \\
                               &+ \sqrt{\int_{|v|\geq c} \int_{|u|\geq c} |\bbW(u,v)-\overline{\bbW}_N(u,v)|^2dudv}
\end{split}
\end{align}
The rightmost term is bounded by $\epsilon(c)$, as $\overline{\bbW}_N$ is zero outside of $[-c,c]$. Since $\bbW$ is $A_w$-Lispchitz in the $[-c,c]$ interval, we can write
\begin{align*}
|\bbW(u,v)-\overline{\bbW}_N(u,v)| &\leq A_w\max\left(\left|u-u_i\right|,\left|u_{i+1}-u\right|\right) \\
&+ A_w\max\left(\left|v-u_j\right|,\left|u_{j+1}-v\right|\right) \\
&\leq {A_w}{\gamma} + {A_w}{\gamma} = {2A_w}{\gamma}
\end{align*}
for $u_i \leq u \leq u_{i+1}$, $u_j \leq v \leq u_{j+1}$, where the $u_i$, $u_j$ are as in Definition \ref{def:graphex} for all $1\leq i,j\leq N$.
Therefore, the leftmost term in \eqref{eqn:integral_term} can be upper bounded as $\sqrt{2c \times 2c \times ({2A_w}{\gamma})^2}=4A_w\gamma c$, which completes the proof.

\subsection{Proof of Theorem \ref{thm:power}}
  
Theorem \ref{thm:power}.1 is a direct consequence of the fact that the graph convolution is pointwise in the spectral domain. To see this, substitute $\bbA=\bbV\bbLam\bbV^\top$ in \eqref{eqn:graph_convolution} and left-multiply both sides by $\bbV^\top$. We get
\begin{equation} \label{eqn:linear_system}
    [\bbV^\top\hby]_i = \sum_{k=0}^{K-1}h_k \lambda_i^k [\bbV^\top\bbx]_i \text{.}
\end{equation}
Hence, Theorem \ref{thm:power}.1 holds for any $h_k \neq 0$.

\noindent
To show Theorem \ref{thm:power}.2, we write \eqref{eqn:graph_convolution} in the matrix form
\begin{equation}
 \hby = [\bbx \  \bbA \bbx  \ldots \  \bbA^{K-1} \bbx] \, [h_0 \ h_1 \ \ldots h_{K-1}]^{\top}. \label{eqn: krylov}  
\end{equation}
To show there exists $h_k$ such that $\hby = \bby$, we consider $K=N$, which yields a linear system of $N$ equations (i.e., $\hby_i = \bby_i$ for $i \in [N]$) with $N$ unknowns $h_0, \ldots, h_{N-1}$. Thus, it suffices to show that the vectors $\bbx, \bbA \bbx, \ldots, \bbA^{N-1} \bbx$ are linearly independent. Consider projecting them to the eigen-basis of $\bbA$, i.e., 
\begin{equation}
    \bbV^{\top} [\bbx \  \bbA \bbx \  \ldots \  \bbA^{N-1} \bbx] \equiv  [\tbx\  \bbLam \tbx\  \ldots\  \bbLam^{N-1} \tbx],
\end{equation}
where $\tbx \coloneqq \bbV^{\top} \bbx$. Since $\bbV$ is invertible, it remains to show that  $\tbx, \ldots, \bbLam^{N-1} \tbx$ are linearly independent. Let $\bm{c} \in \R^N, \bbM \in \R^{N \times d}$, and $\bm{c} \odot \bbM$ denote multiplying the $i$-th row of $\bbM$ by the $i$-th component of $\bm{c}$. We write $\tbx = \bm{c} \odot \bm{1}$ where $\bm{1}$ denotes the all-ones vector. Then the matrix $[\tbx\  \bbLam \tbx\  \ldots\  \bbLam^{N-1} \tbx]$ reduces to 
\begin{equation}
\bm{c} \odot \left[\begin{array}{cccc}
1 & \lambda_1 &  \ldots & \lambda_1^{N-1} \\
1 & \lambda_2 &  \ldots & \lambda_2^{N-1}  \\
\vdots & \vdots &  \ddots & \vdots \\
1 & \lambda_N &  \ldots & \lambda_N^{N-1} 
\end{array}\right]. \label{eqn:vandermonde}
\end{equation}
Observe that \eqref{eqn:vandermonde} is a row-wise scaled Vandermonde matrix, which has determinant $\prod_i [\bm{c}]_i \prod_{i < j} (\lambda_i - \lambda_j)$. By assumption that $[\bbV^{\top} \bbx]_i \neq 0$ for $1 \le i \le N$, all entries $[\bm{c}]_i$ are nonzero. By assumption that the eigenvalues all have multiplicity one, $\lambda_i - \lambda_j \ne 0$ for all $i< j$. Therefore, \eqref{eqn:vandermonde} has nonzero determinant and  linearly independent columns, 
which completes the proof. It is also clear from the proof that both assumptions are necessary for the scaled Vandemonde matrix in \eqref{eqn:vandermonde} to have nonzero determinant.


\subsection{Experiment Details for Sec. \ref{sec:exp-dsgm}} \label{sec:app1}

\noindent \textbf{Data.} Our chosen $\bbW$ in \eqref{eqn:a_graphex} follows the degree-corrected SBM model in Def.\ref{eqn:dc_sbm}, which exhibits block structure via the two parameters $p,q$ and core-periphery pattern via the degree function. It is easy to check that $\bbW$ in \eqref{eqn:a_graphex} satisfies integrability condition in Def. \ref{def:graphex} and the Lipschitz continuity assumption (i) in Thm. \ref{thm:comm_conc}.


\noindent \textbf{Methods.} For a comprehensive investigation, we compare spectral embeddings and GNNs using two graph operators: the graph adjacency matrix $\bbA$  and the normalized adjacency $\tilde{\bbA}$. Since $\bbW$ has $2$ communities, we choose $\phi_{\text{SE}}$ as the top-$K$ eigenvectors of the graph operator where $K \in \{2,6,10,20\}$, combined with the top-$2$ principal components of the nodes features $X$ per \eqref{eqn:node_sp_emb}, and $c_{\text{SE}}$ as a multilayer-perception with 1-hidden layer. We choose $\phi_{\text{GNN}}$ as a degree-2 polynomial graph filter with 2 layers, and $c_{\text{GNN}}$ as a linear layer. All methods are trained with full-batch gradient descent, using learning rate $0.02$ for $200$ epochs (with early stopping if the loss has converged) and dropout probability $0.5$. For GNNs, We use PReLU nonlinearity (i.e., ReLU with a learnable parameter for negative inputs).

\subsection{Experiment Details for Sec. \ref{sec:exp-real}} \label{sec:app2}

\textbf{Data.} The Chameleon webpage network has $2277$ nodes with average node degree $13.8$, where nodes represent webpages and edges are hyperlinks between them. The node features are $2325$-dimensional bag-of-words vectors of the webpages, and node labels are $5$ webpage categories. We use the same data splits (48/32/20 for train/validation/test) from \cite{Pei2020Geom-GCN} released in Pytorch Geometric \cite{Fey2019pytorchgeo}.

\noindent \textbf{Methods.} We use the similar setup as described in Section \ref{sec:app1}, except using SE dimension $K=\kappa \in \{150, 200 \}$, and learning rate $0.01$.

\end{document}